\documentclass[aps,twocolumn,prb,superscript,floatfix,superscriptaddress,showpacs,footinbib]{revtex4-2}

\usepackage{amssymb}

\usepackage[pdftex]{graphicx}
\usepackage{lipsum}
\usepackage{dcolumn}
\usepackage{bm}
\usepackage{amsmath}
\usepackage{array}
\usepackage{color}
\usepackage{subfigure}
\usepackage{dsfont}
\usepackage{txfonts}
\usepackage{multirow}
\usepackage{sidecap}
\usepackage{xcolor,cancel}
\usepackage[normalem]{ulem}

\newcommand{\dn}{\downarrow}

\newcommand{\up}{\uparrow}

\usepackage{comment}

\newcommand{\ket}[1]{{|{#1}\rangle}}

\usepackage{hyperref}
\hypersetup{
    colorlinks,%
    citecolor=blue,%
    linkcolor=blue,%
    urlcolor=blue
}

\begin{document} 

\title{Charge fluctuations and topological phases in Kitaev-Heisenberg ladders}

\author{M. G. Sousa} 

\affiliation{Instituto de F\'isica, Universidade Federal de 
Uberl\^andia, Uberl\^andia, Minas Gerais 38400-902, Brazil.}

\author{O. Ávalos-Ovando}
\affiliation{Department of Nanoscience, Joint School of Nanoscience and Nanoengineering, University of North Carolina at Greensboro, Greensboro, North Carolina 27401, USA}

\author{E. Vernek}
\email{vernek@ufu.br}
\affiliation{Instituto de F\'isica, Universidade Federal de
Uberl\^andia, Uberl\^andia, Minas Gerais 38400-902, Brazil.}

\author{S. E. Ulloa}
\affiliation{Department of Physics and Astronomy and Nanoscale and Quantum Phenomena Institute, Ohio University, Athens, Ohio 45701-2979, USA}


\date{\today}

\begin{abstract}
We investigate the stability of topological phases in doped Kitaev-Heisenberg ladders by studying the competition with itinerant electrons and the associated charge fluctuations in  
a Hubbard model on a honeycomb ribbon geometry.  We analyze the evolution of string order parameters, spin correlations, and charge fluctuations as functions of hopping amplitude and interaction strength in a half-filled band. Our results from density matrix renormalization group (DMRG) calculations show that increasing electron bandwidth progressively suppresses the topological phases, shifting and narrowing their stability regions in the phase diagram. We identify the critical values of hopping where string order vanishes and characterize the interplay between magnetic order and charge fluctuations. These findings provide insight into the robustness of topological phases against doping and charge dynamics, with implications for candidate Kitaev materials and engineered quantum systems.

\end{abstract}
\maketitle

{\it Introduction.} Quantum spin liquids (QSLs) and other topological phases of matter have become pivotal subjects in condensed matter physics due to their nonlocal entanglement properties and potential applications in quantum computation~\cite{Kitaev2006,Savary2016, Knolle2019}. Since its conceptualization by P.~W. Anderson in the famous resonating valence bonds state in triangular lattices~\cite{Anderson1973},  QSLs have been expected to emerge in different systems, including 2D materials. Candidates for QSLs in triangular lattices include organic materials~\cite{PhysRevLett.91.107001,Etme}, rare-earth-based oxides~\cite{PhysRevLett.115.167203}, different frustrated kagome lattices~\cite{PhysRevLett.98.107204,RevModPhys.88.041002, Han2012,Feng2017}, and suitably designed trapped ion arrays~\cite{Friedenauer2008}.
Other prominent candidates for QSLs are materials with a structure tantalizingly close to the famous Kitaev model (KM) on a honeycomb lattice~\cite{Kitaev2006}.  Examples include $\alpha$-RuCl$_3$~\cite{Kasahara2018}, Na$_2$IrO$_3$~\cite{PhysRevB.82.064412},
 H$_3$LiIr$_2$O$_6$~\cite{Kitagawa2018}, and twisted MoTe$_2$~\cite{Mote2}. Magnetically frustrated 3D pyrochlore materials~\cite{PhysRevLett.82.1012, PhysRevX.1.021002} are also under study as possible candidates for QSL behavior. 

The exactly solvable KM on a honeycomb lattice offers a unique platform to study its interesting phase with Majorana excitations and local $Z_2$ gauge fields in great detail~\cite{Kitaev2006}. 
The identification of materials such as $\alpha$-RuCl$_3$ and various iridates as candidates for realizing Kitaev physics, thanks to spin-orbit coupling and characteristic bond-anisotropic interactions, has motivated a great deal of work~\cite{PhysRevB.90.041112,Banerjee2016,Takagi2019}. 
However, the real-world materials inherently include additional interactions, such as an isotropic Heisenberg exchange between neighboring spins.  When Kitaev interactions compete with such magnetic couplings, the resulting Kitaev-Heisenberg model (KHM) no longer admits exact solutions. However, numerical approaches have shown that these systems exhibit rich phase diagrams, including QSLs, magnetically ordered phases, and topologically non-trivial  ground states~\cite{PhysRevLett.105.027204, Rau2016,PhysRevLett.105.027204,PhysRevB.83.245104,PhysRevB.86.224417,PhysRevB.92.020405}. A fascinating aspect of the topological phases of the KM, which remains valid for weak Heisenberg coupling, is that they cannot be described by local order parameters. Instead, the unique topological character needs to be uncovered by nonlocal order parameters, such as the string operator~\cite{PhysRevB.40.4709,Kennedy1992,PhysRevB.45.304,Oshikawa1992,PhysRevB.99.195112}, a quantity that allows challenging but possible experimental observation~\cite{Endres2011}.

Beyond the spin dynamics in Hamiltonians such as the KM and KHM, further complexities may arise in real materials, leading to richer physics. An example of this are possible charge fluctuations that can disrupt a pristine QSL state and stabilize competing ordered phases~\cite{Winter2017}. This phenomenon may arise in $\alpha$-RuCl$_3$ in proximity to graphene, which provides the insulator with itinerant electrons~\cite{PhysRevLett.123.237201}, or in hole-doped Na$_2$IrO$_3$~\cite{PhysRevLett.109.266406, PhysRevLett.111.037205}, and the associated charge fluctuations.  This requires a more general model to allow for charge fluctuations in the KHM or KM descriptions. Perhaps the simplest generalization towards this end is the tunneling Kitaev-Heisenberg model (t-KHM) that introduces an itinerant charge degree of freedom~\cite{PhysRevB.110.224518}. Such a model can be viewed as a Hubbard Hamiltonian competing with the Kitaev ($K$) and Heisenberg ($J$) exchange couplings. In this model, the physics of the pure KHM would be recovered at the particle-hole symmetric point and for strong Coulomb repulsion $U$. Away from particle-hole symmetry, this model has been shown to exhibit hole pairing tendencies in various parameter regimes~\cite{PhysRevB.110.224518}, offering alternative pathways to non-conventional superconductivity~\cite{PhysRevB.86.085145,PhysRevB.85.140510,PhysRevB.87.064508}.  

In this work, we investigate the stability of various quantum phases of Kitaev-Heisenberg ladders under a generalized t-KHM, examining how charge fluctuations and itinerancy influence the exotic quantum orders predicted in the pure-spin KHM. Using density matrix renormalization group (DMRG) calculations, we systematically analyze the evolution of spin correlations and charge fluctuations, as well as string order parameters that signal the onset of topological phases, as functions of hopping amplitude $t$, and different exchange couplings. Our results show that increasing bandwidth in this system progressively limits the overall amplitude of the nonlocal string operators, shifting the parameter range over which topological phases appear in this model.


\begin{figure}[t!]
    \centering
    \subfigure{\includegraphics[clip, width=0.475\textwidth]{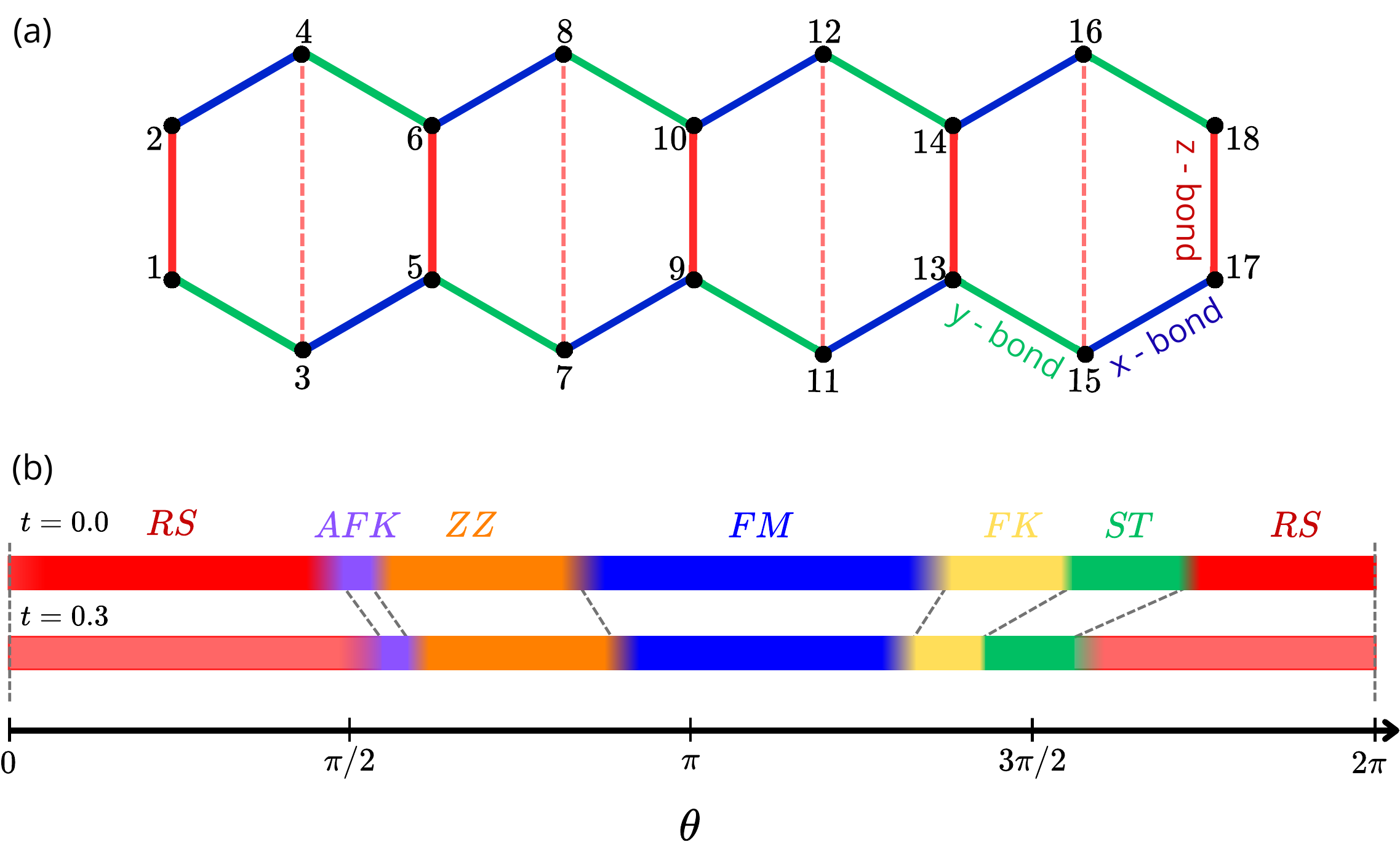}} 
    \caption{(a) Schematic representation of the honeycomb ribbon with four rings, showing the bonds associated with the anisotropic Kitaev model with color-coded bonds. Dashed red lines indicate bonds across the two legs of the ladder. (b) Sketch of the phase diagram of the system governed by the extended Heisenberg-Kitaev model, t-HKM, for $t=0$ (top) and $t=0.3$ (bottom).  The $\theta$ axis shows shifts in the phases as $t$ increases.  The value of $\theta$ determines the exchange constants $J=A \cos \theta$ and $K=A \sin \theta$, where A is the total exchange coupling (see main text).  The color bar regimes indicate different known phases. FM: ferromagnetic; FK: ferromagnetic Kitaev; ST: stripy or FM-rung Ising; RS: rung singlet;  AFK: antiferromagnetic Kitaev;  ZZ: FM-leg Ising, where rungs are vertical $z$-bonds and legs are $x,y$ bonds in (a).}
    \label{figmodel}
\end{figure}
%

%

{\it Model and methods.} Our system is inspired by doped RuCl$_3$, and consists of a honeycomb ribbon, as illustrated in Fig.~\ref{figmodel}. As described above, the system is modeled by Heisenberg-Kitaev couplings supplemented by a Hubbard Hamiltonian. This can be written as 
\begin{eqnarray}\label{hamilton}
	H &=& H_{\rm Hubb} + H_{\rm H-K} \, .
	\end{eqnarray}
Here, 
\begin{equation}
	H_{\rm Hubb} =\varepsilon_0\sum_{i\sigma } n_{i\sigma} -t \sum_{\langle i,j \rangle, \sigma} \left( c_{i,\sigma}^\dagger c_{j,\sigma} + \text{H.c.} \right) + U \sum_{i} n_{i,\up} n_{i,\dn}
\label{hubbard}
\end{equation}
is the Hubbard  Hamiltonian describing the electronic degrees of freedom, with
$c_{i,\sigma}^\dagger$ ($c_{i,\sigma}$) representing the creation (annihilation) operator for electrons with energy $\varepsilon_0$ and spin $\sigma$ at site $i$, $n_{i,\sigma} = c_{i,\sigma}^\dagger c_{i,\sigma}$ is the number operator. The parameter $t$ is the nearest-neighbor hopping amplitude, and $U$ is the on-site Coulomb repulsion.  
The Heisenberg-Kitaev model is given by
\begin{eqnarray}
	\label{HK}
	H_{\rm H-K} = J\sum_{\langle i,j \rangle} \, \mathbf{S}_{i} \cdot \mathbf{S}_{j} + K\sum_{\langle i,j \rangle_c} \, S_{i}^c S_{j}^c \, ,
\end{eqnarray}
which describes the magnetic interactions between the localized spins  $\mathbf{S}_i = \frac{1}{2} \sum_{\alpha,\beta} c_{i,\alpha}^\dagger \, \boldsymbol{\sigma}_{\alpha\beta} \, c_{i,\beta}$  at each site $i$ of the honeycomb lattice, with  $\boldsymbol{\sigma} = (\sigma^x, \sigma^y, \sigma^z)$  being the Pauli matrices, and $\alpha, \beta \in \{\uparrow, \downarrow\}$ denote spin indices $(\hbar=1)$. 
The first term corresponds to the isotropic Heisenberg nearest-neighbor coupling $J$, which promotes magnetic ordering, while the second one is the anisotropic Kitaev  $K$ that selectively couples the $S^c$ spin components along the specific bond direction, where $c \in \{x, y, z\}$, as indicated by the different colors on the ladder (Fig.~\ref{figmodel}a).  For concreteness, we set  $\varepsilon_0=-U/2$, ensuring the particle-hole symmetry of the Hubbard Hamiltonian.

When $ t=0$ and $ U=-2\varepsilon_0 > 0$, the charge degrees of freedom are frozen and the Hamiltonian \eqref{hamilton} captures the physics of the complete Kitaev-Heisenberg spin model. As mentioned above, in this case, the KHM exhibits a rich phase diagram, including conventional magnetic order and topological phases, depending on the values of $K$ and $J$. The topological phases are characterized by the presence of nonlocal string order parameters (SOPs) that reveal long-range spin correlations. One suitable string order operator can be defined for a spin string of length $r$, as identified in \cite{PhysRevB.99.195112},    
\begin{equation}
O^z(r) = 
\begin{cases}
    (-1)^p \biggl\langle \sigma_1^y \sigma_2^x \left(\prod_{k=3}^{2p} \sigma^z_k\right) \sigma^y_{2p+1} \sigma^x_{2p+2} \biggr\rangle & \text{for $r$ odd}, \\
    (-1)^p \biggl\langle \sigma_1^y \sigma_2^x \left(\prod_{k=3}^{2p} \sigma^z_k\right) \sigma^x_{2p+1} \sigma^y_{2p+2} \biggr\rangle & \text{for  $r$ even}. 
\end{cases}
\end{equation}     
Here, $p = \text{int}\left\lfloor \frac{r - 1}{2} \right\rfloor + 1$ and the $\langle \cdots \rangle$ represents the expectation value taken over the ground state.
The alternating structure of $O^z(r)$ reflects the bond-dependent coupling of the Kitaev ladder model and ensures the correct reconstruction of spin correlations~\cite{PhysRevB.99.195112}. Our focus here is to study how the presence of a finite hopping amplitude  $t$ modifies the different phases in the KHM, by monitoring the stability of this SOP as $t$ increases. 
It is convenient to evaluate the SOP for a given string length $r_0$ as a representative value of the SOP in a system of total length $L$, defined as 
\begin{equation}
	\mathcal{O}_0^z = \sqrt{ \left| O^z(r_0) \right| } \, , 
\end{equation}
where the choice $r_0 = 3L/4$ reduces effects of the open end boundary, ensuring that $\mathcal{O}_0^z$ captures the behavior in the thermodynamic limit. A finite value of $\mathcal{O}_0^z$ as $L \to \infty$ indicates the presence of topological order, characterizing the AFK and FK phases \cite{PhysRevB.99.195112}.

We employ the DMRG to compute the ground state of the full Hamiltonian, from which the SOP is obtained and analyzed as the model parameters $J$, $K$, and $t$ change. Other correlations and local properties are also computed to characterize the phases of the system, including spin correlations and charge fluctuations. The DMRG calculations are performed using the ITensor library~\cite{itensor,itensor-r0.3}, a suitable tensor network platform that enables us to efficiently handle the inherent complexity of the correlated eigenstates of the Hamiltonian.

\label{results}
{\it Numerical results.} To obtain our numerical results, we consider a system with $N$ hexagons containing $4N+2=L$ sites as shown in Fig.~\ref{figmodel}. For simplicity, we  parameterize $J = A\cos\theta$ and $K = A \sin\theta$, with $\theta \in [0, 2\pi]$, and measure energy in units of the total exchange $\sqrt{J^2 + K^2}=A$. This allows us to explore the full phase diagram of the Kitaev-Heisenberg model by varying $\theta$.  We consider a large value of $U=10$, which would be expected to result in weak charge fluctuations in the system, provided $t \ll U$. We present results for ribbons with $N=10$ rings, which are well handled by DMRG calculations while being large enough to manifest the system's phases in the thermodynamic limit.

\paragraph{Stability of the FK phase.---} 
\begin{figure}[b!]
	\centering
	\subfigure{\includegraphics[clip,width=3.40in]{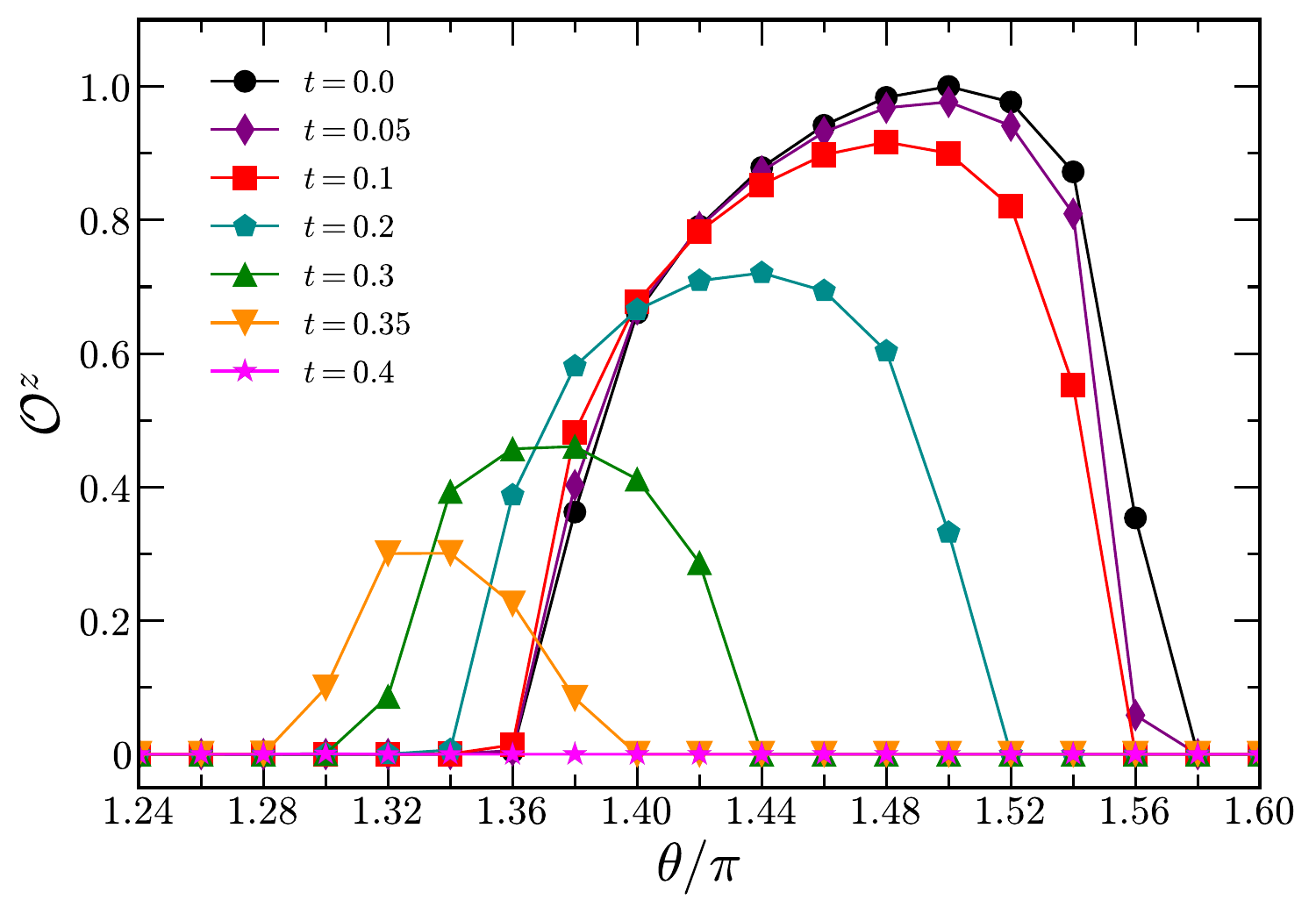}}
	\caption{String order parameter ${\cal O}_0^z$ as function of $\theta$ near the FK phase for various values of  $t$. Note the maximum value of ${\cal O}_0^z$ is attained at $\theta=3\pi/2$ for $t=0$ (black circles), as expected $(K=-1, J=0)$. Finite $t$ suppresses the maximum and shifts the region of nonzero SOP towards lower  values of $\theta$.  The SOP eventually vanishes for $t\ge0.4$.}
	\label{fig2}
\end{figure}
We start by investigating the stability of the ferromagnetic Kitaev phase (FK) that exists over a relatively wide range around $\theta=3\pi/2$. This parameter regime results in a QSL phase in the two-dimensional (2D) honeycomb lattice \cite{PhysRevLett.105.027204}.  Figure~\ref{fig2} shows the SOP as a function of $\theta$  for different values of $t$ in the region $\theta \simeq 3\pi/2$. For $t=0$ (black circles) the SOP shows a pronounced enhancement extending from $\theta \approx 1.36 \pi$ to $\theta \approx 1.57 \pi $, reaching a maximum value near unity at $\theta = 3\pi/2$. This behavior is in full agreement with Fig.~3(b) in \cite{PhysRevB.99.195112}. As $t$ increases, there is a progressive suppression of the SOP maximum value and a shift of the entire region with nonzero SOP towards lower values of $\theta$.
The shift to lower $\theta$ can be seen as a weakening of the FM phase relative to the FK phase, despite a small Heisenberg coupling $J~( < 0)$, while the topological character remains sizable near $K \approx -1$. The width of the nonzero SOP FK phase also decreases with increasing $t$. These trends suggest that systematically increasing $t$ will eventually cause the FK phase to disappear. Indeed, the SOP is seen to vanish for $t\ge 0.4$. We return to this point below.

\paragraph{Stability of the AFK phase.---} 
\begin{figure}[t!]
	\centering
	\subfigure{\includegraphics[clip,width=3.40in]{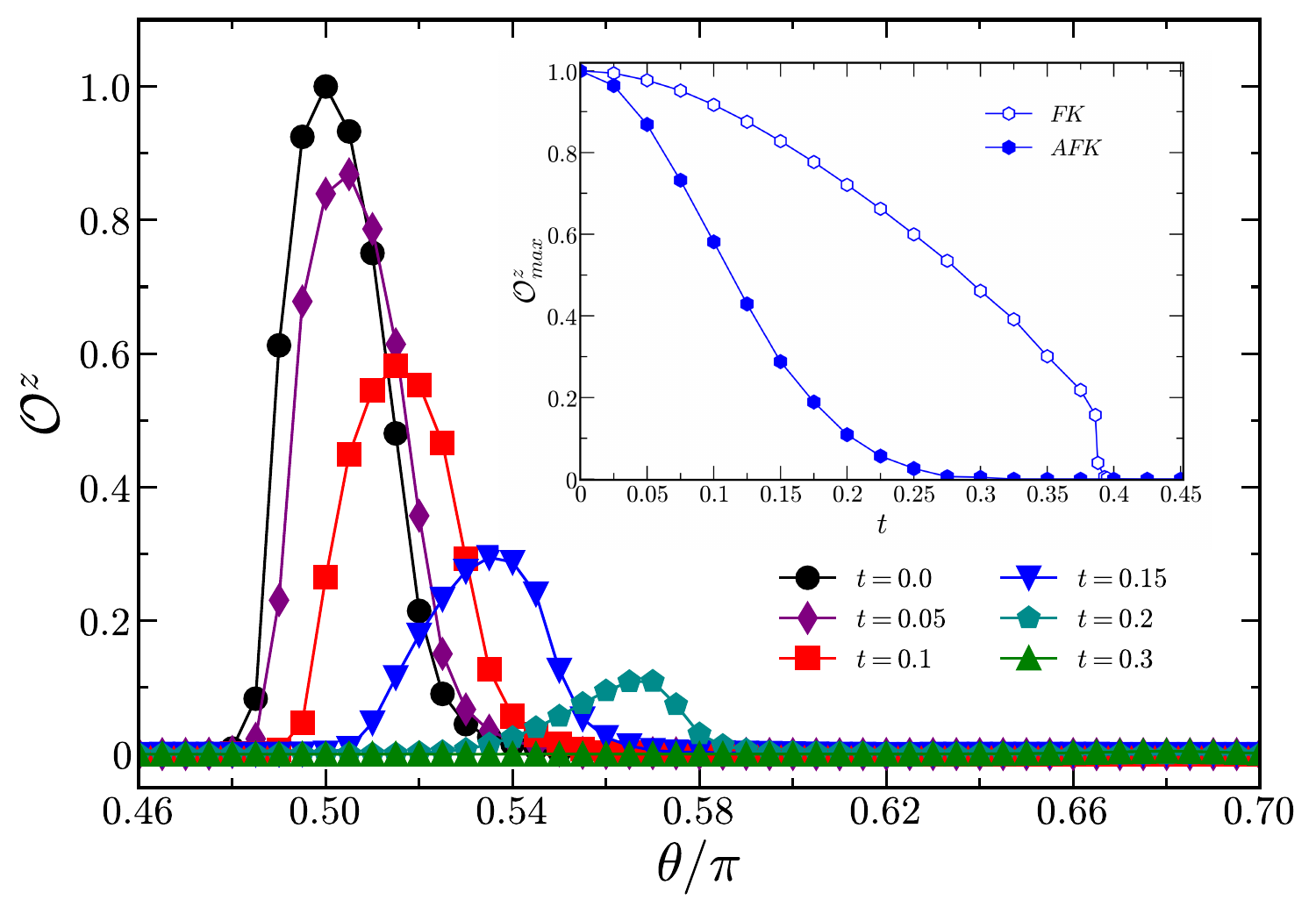}}
	\caption{String order parameter ${\cal O}_0^z$ as function of $\theta$ near the AFK phase for various values of  $t$. The maximum value of ${\cal O}_0^z=1$ at $\theta=\pi/2$ for $t=0$ (black circles), identifies the optimal AFK phase. Finite $t$ suppresses and shifts of SOP towards larger values of $\theta$, and eventually vanishes for $t \gtrsim 0.3$. Inset: Maximum value of $\mathcal{O}^z$ as a function of $t$ for the FK (open hexagons) and AFK (filled hexagon) phases. Maximum values of the SOP extracted from Fig.~\ref{fig2} and Fig.~\ref{fig3}. Notice that in both cases the $\theta$ for maxima depends on $t$. \label{fig4}}
	\label{fig3}
\end{figure}
Let us now analyze how $t$ affects the stability of the AFK phase. This QSL phase in the 2D honeycomb lattice is present near $\theta \simeq \pi/2$ in the KHM. 
Figure~\ref{fig3} shows the SOP as a function of $\theta$ in this region for different values of $t$. For $t=0$ (black circles), we observe a pronounced peak with a maximum of 1 at $\theta=\pi/2$, which corresponds to the perfect AFK phase~\cite{PhysRevB.99.195112}.
As $t$ increases, there is a progressive suppression of the SOP and a shift of the peak towards larger values of $\theta$.  
The AFK is evidently more fragile for finite $t$ than the FK phase. We see that for $t=0.3$ the SOP has already vanished in the AFK, while it remains noticeably finite for $t \approx 0.35$ in the FK phase. This difference in the stability of the two phases arises from the nearest spins having opposite correlation in the AFK phase, so that charge fluctuations induced by a finite $t$ are more effective.  In the FK phase, the spins are aligned, which helps suppress charge fluctuations through Pauli's exclusion principle.

%

To further compare the effect of $t$ in the AFK and FK phases, we plot the maximum value of $\mathcal{O}_0^z$ as function of $t$ for these two phases. The results are shown in the inset of Fig.~\ref{fig4} for the FK (open hexagon) and AFK (filled hexagon) phase. The maximum value of $\mathcal{O}_0^z$ decreases in both phases as $t$ increases, but the drop is faster in the AFK than in the FK phase. For example, while $(\mathcal{O}_0^z)_{\rm max}$  has already vanished for $  t\approx 0.3$ in the AFK, it is still finite in the FK phase.

\paragraph{Charge and local momentum changes.---}
A natural question is the reason for the changes in SOP and the associated phases.
As one would anticipate, this suppression in SOP results from increasing local charge fluctuation arising from the hopping degree of freedom.  Hopping-aided processes are expected to reduce the overall spin at each site, thereby weakening the magnetic order induced by the Heisenberg and Kitaev terms.

\begin{figure}[t!]
	\centering
	\subfigure{\includegraphics[clip,width=3.40in]{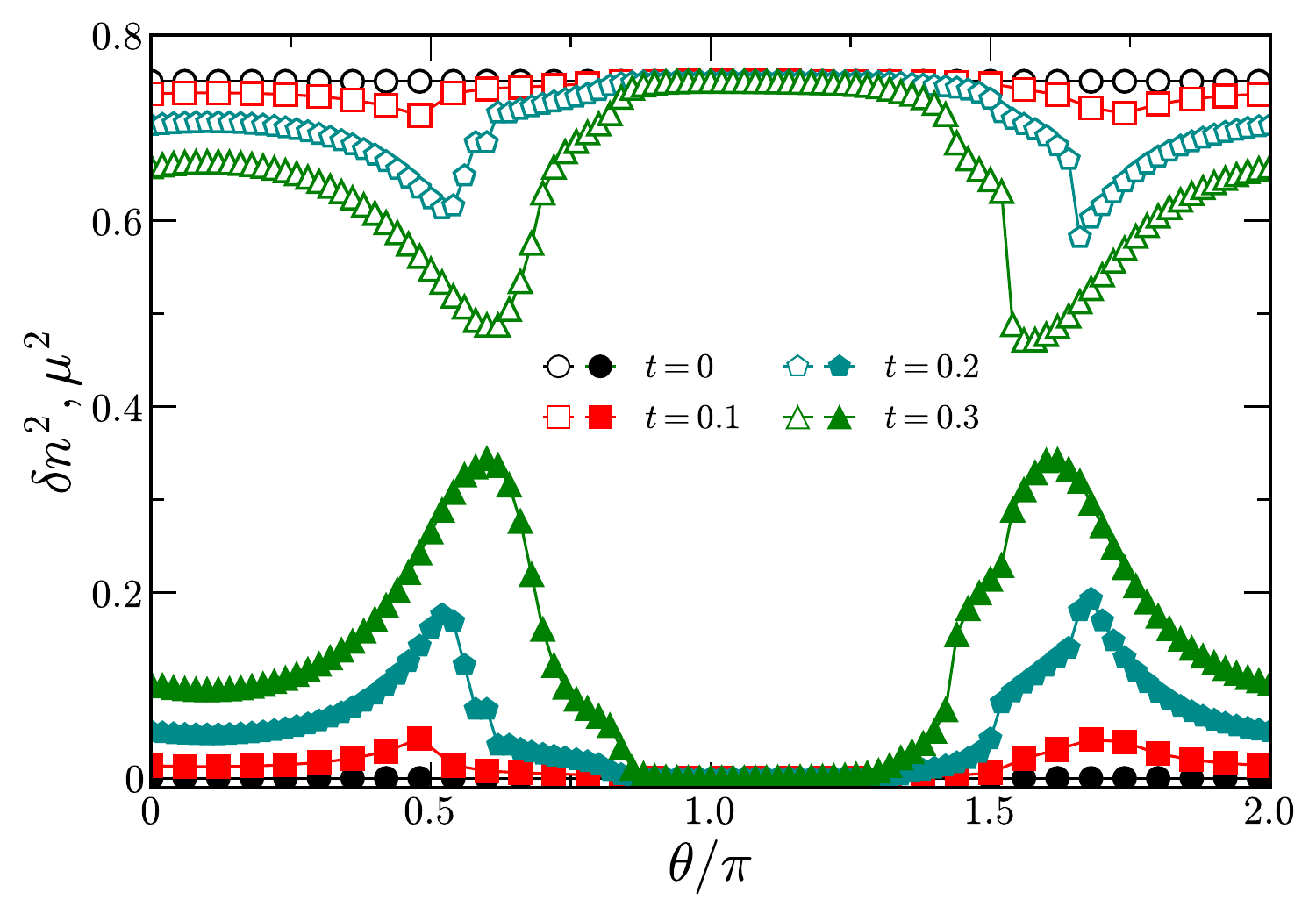}}
	\caption{Local magnetic moments, $\mu^2$, (open symbols) and charge fluctuations, $\delta^2$, (filled symbols) as a function of $\theta$ for various values of the hopping $t$. For t = 0 there are no charge fluctuations while the magnetic moment is close to $3/4$, as expected for a full spin $1/2$ per site. As t
increases, two charge fluctuations peaks emerge near the AFK and FK phases. These peaks are accompanied by dips in the magnetic moment in the same positions
}
	\label{fig6}
\end{figure}

For a quantitative assessment of the effect of the hopping $t$ in the different phases, we calculate the local charge fluctuation  
\begin{eqnarray}
(\Delta n_i)^2 &=&  \langle n_i^2 \rangle -\langle n_i \rangle^2 \nonumber\\
& =& \langle n_{i\uparrow} \rangle + \langle n_{i\downarrow} \rangle + 2 \langle n_{i\uparrow} n_{i\downarrow} \rangle -(\langle n_{i\up}\rangle + \langle n_{i\dn}\rangle)^2.
\end{eqnarray}
In the half-filled regime, where the system is particle-hole symmetric, we have $\langle n_{i\uparrow} \rangle = \langle n_{i\downarrow} \rangle = 1/2$, and the local charge fluctuation reduces to $(\Delta n_i)^2 = 2 \langle n_{i\uparrow} n_{i\downarrow} \rangle$. 
At half-filling, then, $(\Delta n_i)^2$ is proportional to the double occupancy $\langle n_{i\uparrow} n_{i\downarrow} \rangle$, associated with the probability of finding two electrons with opposite spins occupying the same site. Thus, an enhancement of $(\Delta n_i)^2$  should suppress local moments, which can be extracted from the diagonal elements of the correlation function between sites $i$ and $j$,
\begin{equation}
	C_{i,j} = \langle \mathbf{S}_i \cdot \mathbf{S}_j \rangle , 
\end{equation}
where the expectation value is taken over the ground state. We define the average local moment as $\mu^2=\sum_i C_{i,i}/L$, where $L$ is the number of sites in the system. The average reduces the potential non-uniformity caused by the system's boundaries. Similarly, we define the average charge fluctuation as $\delta n^2=\sum_i (\Delta n_i)^2/L$. Figure~\ref{fig6} shows $\delta n^2$ (open symbols)  and $\mu^2$ (closed symbols) as a function of $\theta$ for different values of $t$. We observe that for $t=0$ (black circles) the local charge fluctuation vanishes for all $\theta$, while the local moment remains at $\mu^2 =3/4$, as it corresponds to a well-defined spin $1/2$ at each site. As $t$ increases, we observe a progressive increase in $(\delta n)^2$ accompanied by a suppression of $\mu^2$, with the changes being more pronounced near the AFK ($\theta/\pi \simeq 1/2$) and FK ($\theta/\pi \simeq 3/2$) phases. This is clearly visible in the two pronounced dips in $\mu^2$ and peaks in $\delta n ^2$ for $t=0.3$ (red symbols).  It is interesting to note that the $\theta$ shifts described in Fig.~\ref{fig2} and \ref{fig3} as $t$ increases occur towards the FM phase, reducing the $\theta$ range over which such a phase exists.  That is evident in Fig.~\ref{fig6} where the peaks in $\delta n^2$ (or dips in $\mu^2$) move toward each other as $t$ increases. 
It must also be pointed out that the range over which the SOP is nonzero near the FK and AFK phases is also reduced with increasing $t$, as illustrated in Fig.~\ref{figmodel}, and accompanied by an overall enhancement of the charge fluctuations and drop in local moment.  An additional point of note is that the FM phase centered around $\theta\simeq\pi$ is clearly more robust to changes in $t$, suppressing charge fluctuations and maintaining spin 1/2 local moments even for large $t$ value (see Fig.~\ref{fig6}).

\begin{figure}[h]
	\centering
	\subfigure{\includegraphics[clip,width=3.40in]{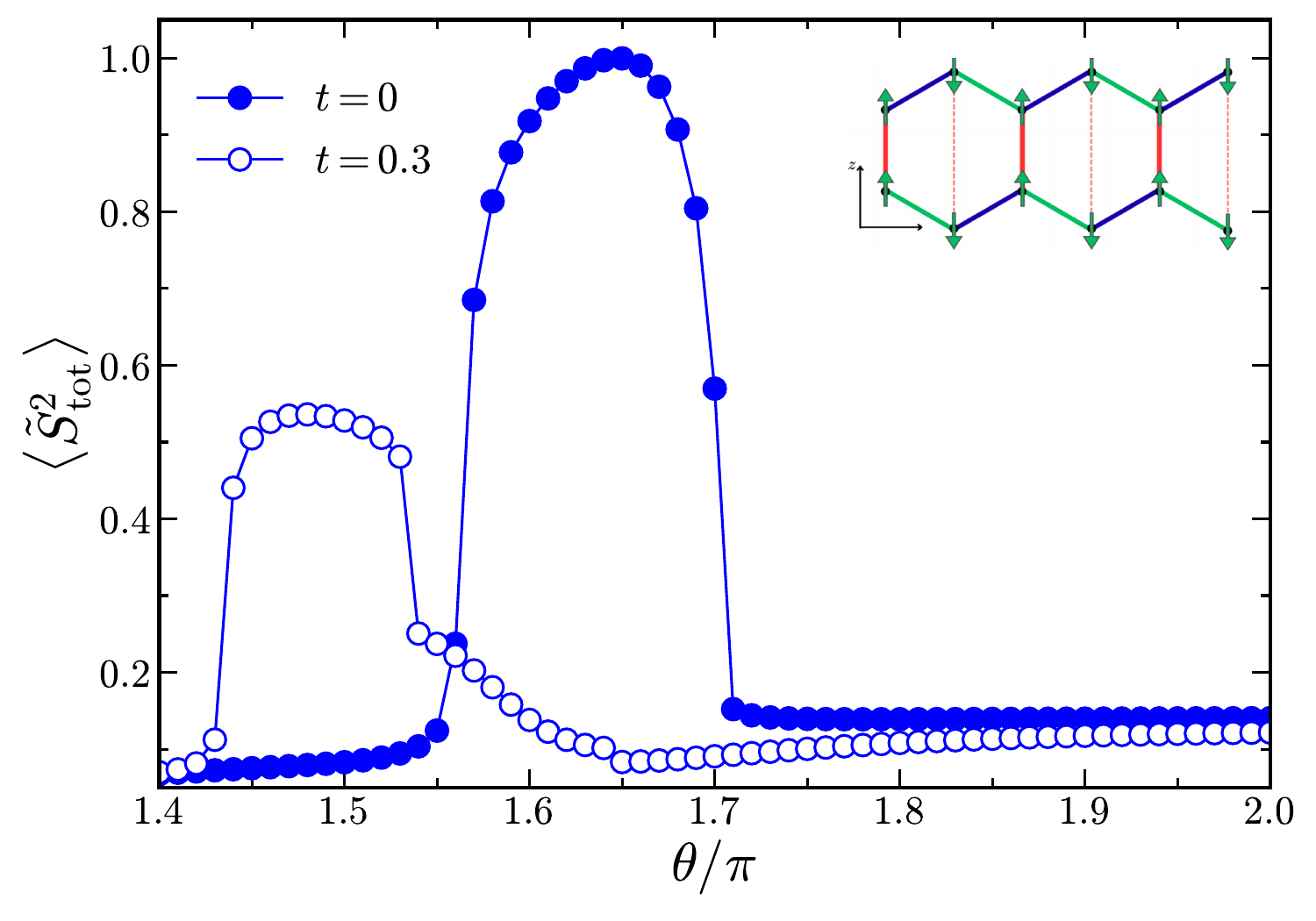}}
	\caption{Stripy rotated order parameter, $\langle \tilde{S}_{\text{tot}}^2 \rangle$, as a function of $\theta$. Filled blue circles correspond to $t=0$, indicating the presence of the stripy phase in the range $\theta/\pi \in [1.56, 1.71]$. Open circles correspond to $t=0.3$, indicating that the ST phase has shifted to the range $\theta/\pi \in [1.43, 1.54]$. The inset shows a schematic representation of the spin arrangements in the ladder in the stripy phase.}
	\label{fig5}
\end{figure}
\paragraph{Stability of the stripy phase.--- }
The stripy phase ground state is characterized by the spins being aligned ferromagnetically along one lattice direction and antiferromagnetically along the others, thus forming a stripy pattern. The order can be seen as ferromagnetic (FM) alignment along the vertical rungs and antiferromagnetic (AFM) alignment along the ladder legs (see inset, Fig.~\ref{fig5}). The phase has an order parameter based on the total spin moment evaluated on a modified rotated basis that transforms the stripy pattern into an effective ferromagnetic state. Following~\cite{PhysRevLett.105.027204}, we  define a four-site sublattice that imparts rotations to each spin component of the neighbors according to their index
\begin{eqnarray}\label{stripy_OP}
\langle \tilde S_{\mathrm{tot}}^2 \rangle
= \frac{1}{{\cal N}_L} \sum_{i,j=1}^L\sum_{\alpha }  \sum_{p^\alpha\in \{A,B,C,D\} } p_i^\alpha p_j^\alpha \,
\langle S_i^\alpha S_j^\alpha \rangle,
\end{eqnarray} 
where the coefficients $p_i^\alpha$ and $p_j^\alpha$ take values in the triple sets $A=(+1,+1,+1),\ B=(-1,-1,+1),\  C=(-1,+1,-1)$ and $ D=(+1,-1,-1)$ that accompany the spin operators $S_i^\alpha$ and $S_j^\alpha$. These sets define sublattices to make the stripy phase effectively ferromagnetic. The normalization factor in Eq.~\eqref{stripy_OP} is given by ${\cal N}_L=\tilde S_{\rm tot}(\tilde S_{\rm tot}+1)$, which corresponds to the expectation value of the effective total spin squared, $\tilde S_{\rm tot}^2$, where $\tilde S_{\rm tot}=L/2$ in the effective ferromagnetic stripy phase.

Figure \ref{fig5} shows $\tilde S_{\rm tot}^2$ vs $\theta$ in the range
$1.4 \pi \leq \theta \leq 2\pi$ for $t=0$ (closed circles) and $t=0.3$ (open circles). We note that $\tilde S_{\rm tot}^2$ is sizable only in a narrow region, in agreement with previous work \cite{PhysRevLett.105.027204, PhysRevB.99.195112} reaching unity at  $\theta \approx 1.65\pi$ for $t=0$. For $t=0.3$, the values of $\tilde S_{\rm tot}^2$  are suppressed and the region of nonzero values moves to lower values of $\theta$, in agreement with the shifts seen for the neighboring FK phase.  Notice that over the entire stripy ST phase, charge fluctuations are sizeable, and the local magnetic moments are suppressed for nonzero $t$.  It is also interesting to notice in Fig.~\ref{fig5} that the transition from the stripy ST phase to the rung singlet RS phase, which is sharply identified for $t=0$ at $\theta /\pi \simeq 1.71$, has become less well-defined for $t=0.3$.

\begin{figure}[t!]
    \centering
    \subfigure{\includegraphics[clip,width=3.2in]{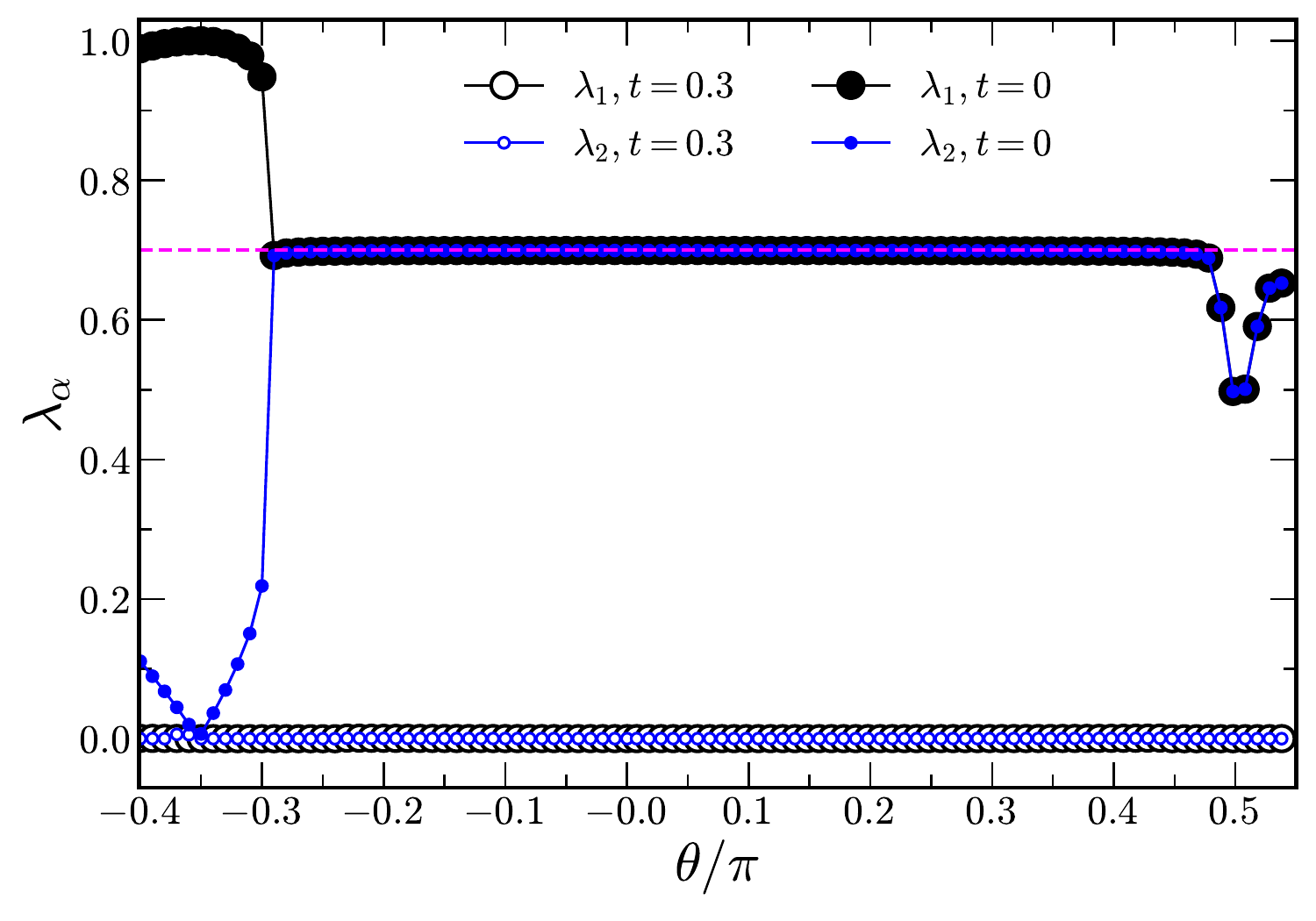}}
    \caption{Schmidt coefficient $\lambda_1$ (black circles) and $\lambda_2$ (blue) vs $\theta$ showing the rung singlet phase. Closed and open symbols correspond to $t=0$ and $t=0.3$. The dashed magenta line indicates the theoretical expected value for a singlet,  $\lambda_\alpha =1/\sqrt{2} \approx 0.707$. Note that for $t=0$ the largest Schmidt coefficients are doubly degenerate in the region of the RS rung singlet phase, $-0.3\pi \leq \theta \lesssim 0.48\pi$. For $t=0.3$, the signature of this phase is completely suppressed.}  
    \label{fig7}
\end{figure}

\paragraph{Stability of the rung-singlet phase.---}
Let us now look at the effect of hopping $t$ on the rung singlet RS phase stabilized for $\theta \simeq 0$. This phase corresponds to singlets formed by pairs of spins connected by $z$-bonds, and a proper way to characterize it is via entanglement entropy. The Schmidt coefficients can be used as a measure of the entanglement along the $z$-bonds~\cite{PhysRevB.99.195112}. To this end, we partition the system into two leg subsystems, A and B, connected by a given $z$-bond, so that 
a given eigenstate can be decomposed as $\ket{\psi}=\sum_{\alpha=1}^M \lambda_\alpha \ket{\psi_\alpha}_A\otimes\ket{\psi_\alpha}_B$, where $0\leq \lambda_\alpha \leq 1$ are the Schmidt coefficients and $M$ is the bond dimension. The entanglement entropy is then given by $S_E=-\sum_\alpha \lambda_\alpha^2\ln \lambda_\alpha^2$. We average the Schmidt spectrum over all $z$-bonds to minimize the effects of open boundary conditions. Figure~\ref{fig7} shows the two largest Schmidt coefficients extracted from the ground state for $-0.4\pi \leq \theta \leq 0.55 \pi $,  comprising the rung singlet phase. As expected, for $t=0$ the maximum value of the Schmidt spectrum is doubly degenerate ($\lambda_1=\lambda_2=1/\sqrt{2}$, solid symbols) in the rung singlet phase, which exists within $-0.3 \pi \lesssim \theta \lesssim 0.48 \pi$~\cite{PhysRevB.99.195112}.  These $\lambda_i$ values indicate strong entanglement between the two spins connected by the $z$-bonds.  In contrast, for $t=0.3$ (open symbols), we note that the largest Schmidt coefficients decrease dramatically (as for increasing $t$ more coefficients contribute to the normalized spectrum, $\sum_\alpha \lambda_\alpha^2=1$). This drop for larger $t$ indicates that the rung singlets are strongly suppressed.  

\begin{figure}[!htbp]
\centering
\subfigure{\includegraphics[clip,width=3.3in]{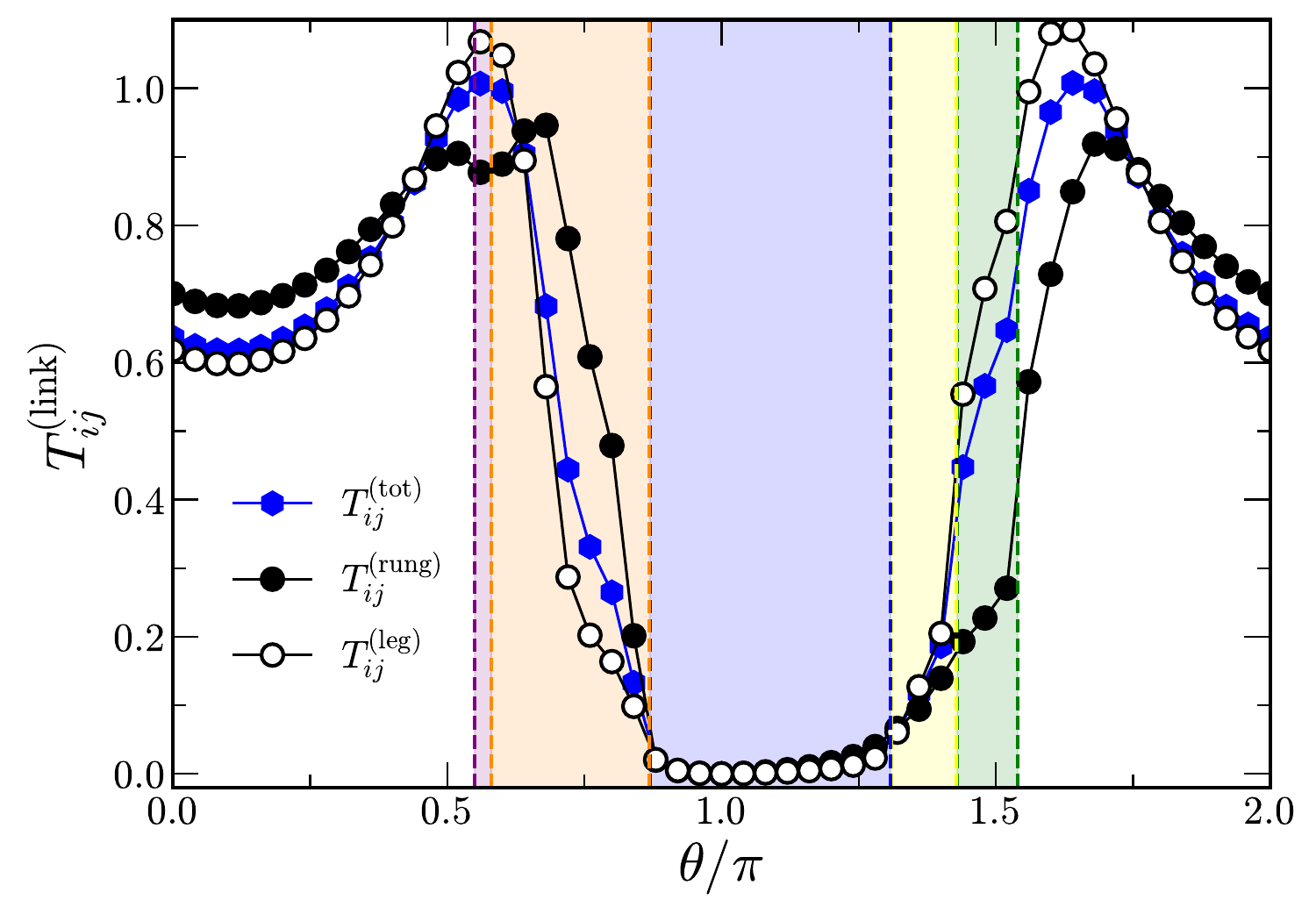}}
\caption{Kinetic bond-correlator $T^{\rm link}_{ij}$ for different link directions as function of $\theta/\pi$, for $t=0.3$. The color bands correspond to the range of the different phases of this model. Each quantity is averaged over all bonds of the same type to minimize boundary effects.} 
\label{fignew}
\end{figure}

\paragraph{Charge mobility}
As carrier hopping is allowed in the system, it competes with the spin correlation energetics.  To quantify the electronic mobility, we introduce the kinetic bond-correlator, the expectation value of the kinetic operator, $T_{ij}^{\rm link} = 2\,{\rm Re} \sum_{\sigma } \langle c_{i\sigma}^\dagger c_{j\sigma}\rangle _{\rm link}$.  This measure of electron delocalization would be expected to depend on the bond-ordering direction, motivating us to explore it for different bonds: rung ($z$–bonds) and leg links ($x$ and $y$–bonds), giving an insight into how electronic mobility depends on the underlying magnetic order. Notice the kinetic correlator is identically zero in the absence of hopping for any of the phases. 
Figure~\ref{fignew} illustrates the kinetic bond-correlator as a function of $\theta$ for $t=0.3$. 
$T_{ij}^{\rm leg}$ includes only the $i,j$ pairs corresponding to the $x,y$ bonds (blue, green in Fig.~\ref{figmodel}) along one leg of the ladder, while $T_{ij}^{\rm rung}$ 
considers $z$-bonds (black circles); $T_{ij}^{\rm tot}$ includes all bonds. For all cases we average over all bonds of the same type to minimize boundary effects. 
The FM phase (blue shadding, $\theta \simeq \pi$) exhibits essentially vanishing mobility for all links, as a direct consequence of the strong ferromagnetic alignment that suppresses
charge fluctuations, even as $t\neq 0$. The topological phases of the model, in
contrast, display markedly different behaviors. The FK phase
(yellow) shows relatively low values of $T_{ij}$, reflecting the
dominance of ferromagnetic interactions, whereas the AFK phase ($\theta\simeq \pi/2$), characterized by antiferromagnetic correlations, exhibits higher charge mobility in all 
directions.
A particularly interesting behavior occurs in the ZZ (orange) and ST (green) phases, whose magnetic structures are complementary. In the ZZ phase, spins align ferromagnetically along the legs and antiferromagnetically on the rungs,
while in the ST phase the opposite pattern occurs. Consequently, larger charge fluctuations are seen along the directions
with antiferromagnetic correlations.  Most interestingly,
two pronounced mobility peaks can be
identified near the edges of topological phases.  For $\theta \simeq 0.6\pi$, the kinetic correlators peak at the edge of the AFK phase, while for $\theta \simeq 1.7\pi$ (in the RS phase for $t = 0$) the peak is beyond the ST phase, perhaps indicating an underlying different phase. It is also interesting that in both peaks, the mobility is systematically stronger along the legs of the ladder.

%
{\it Conclusions.} In summary, we have systematically investigated the stability of topological and magnetically ordered phases in doped Kitaev-Heisenberg ladders by employing a Hubbard model with additional Kitaev and Heisenberg interactions. Through extensive DMRG simulations, we analyzed the evolution of string order parameters, spin correlations, and charge fluctuations as a function of the hopping amplitude $t$ and interaction strengths $K$ and $J$. Our results demonstrate that increasing $t$ progressively suppresses the string order, leading to the destabilization and eventual disappearance of both ferromagnetic and antiferromagnetic Kitaev phases. Notably, the antiferromagnetic Kitaev phase is more sensitive to charge fluctuations, losing its topological character at lower $t$ compared to the ferromagnetic phase, which we attribute to differences in spin alignment and the effects of Pauli exclusion. 
We also examined the fate of the stripy and rung singlet phases, finding that their characteristic order parameters and entanglement signatures are diminished as hopping increases. This indicates a reduction of their stability regions in the phase diagram. The interplay between local magnetic moments arising from strong repulsive interactions, magnetic exchange, and itinerant dynamics is found to play a crucial role in determining the robustness of the different quantum phases. Our findings provide valuable insight into how charge dynamics and doping can erode or modify topological and magnetic orders in candidate Kitaev materials and engineered quantum systems. This has direct relevance to experimental efforts on materials such as proximitized $\alpha$-RuCl$_3$, MoTe$_2$, and iridates, as well as in the design of artificial quantum platforms where hopping and interactions can be tuned. Future work could explore the effects of longer-range interactions, doping away from half-filling, disorder, or dynamical properties, further deepening our understanding of the interplay between itinerancy and topological order in strongly correlated systems.

\begin{acknowledgments}
The  authors acknowledge financial support  from CAPES, FAPEMIG and CNPq. EV thanks  CNPq (Process 301714/2025-8). SEU acknowledges support by the US Department of Energy, Office of Basic Energy Sciences, Materials Science and Engineering Division. This  work used computational  resources of the ``Centro Nacional de Processamento de Alto Desempenho em São Paulo (CENAPAD-SP)''.
\end{acknowledgments}

\section*{Data availabitlity}
The numerical data used in this article are not publicly available but may be provided upon reasonable request.

\bibliography{references}

\end{document}